**Title**: Complete reversal of the atomic unquenched orbital moment by a single electron

**Authors**: R. Rejali,[1]* D. Coffey,[1] J. Gobeil,[1] J. W. González,[2] F. Delgado,[3] A. F. Otte[1]†

**Affiliations**

[1] Department of Quantum Nanoscience, Kavli Institute of Nanoscience, Delft University of Technology, Lorentzweg 1, 2628 CJ Delft, The Netherlands.

[2] Departamento de Física, Universidad Técnica Federico Santa María, Casilla Postal 110V, Valparaíso, Chile.

[3] Departamento de Física & Instituto de Estudios Avanzados (IUdEA), Universidad de La Laguna, 38200, San Cristóbal de La Laguna, Santa Cruz de Tenerife, Spain

*r.rejali@tudelft.nl

†a.f.otte@tudelft.nl

**Abstract**

The orbital angular moment of magnetic atoms adsorbed on surfaces is often quenched as a result of an anisotropic crystal field. Due to spin-orbit coupling, what remains of the orbital moment typically delineates the orientation of the electron spin. These two effects limit the scope of information processing based on these atoms to essentially only one magnetic degree of freedom: the spin. In this work, we gain independent access to both the spin and orbital degrees of freedom of a single atom, inciting and probing excitations of each moment. By coordinating a single Fe atom atop the nitrogen site of the $Cu_2N$ lattice, we realize a single-atom system with a large zero-field splitting—the largest reported for Fe atoms on surfaces—and an unquenched uniaxial orbital moment that closely approaches the free-atom value. We demonstrate a full reversal of the orbital moment through a single-electron tunneling event between the tip and Fe atom, a process that is mediated by a charged virtual state and leaves the spin unchanged. These results, which we corroborate using density functional theory and first-principles multiplet calculations, demonstrate independent control over the spin and orbital degrees of freedom in a single-atom system.

## Introduction

Efforts to downscale information storage to the single-atom limit have largely focused on readily probing, manipulating, and engineering the spin of magnetic atoms adsorbed on surfaces.[1–4] This is primarily a consequence of orbital quenching: the orbital angular momentum $\mathbf{L}$ of these systems is often diminished due to the interaction between the spin-orbit coupling and the crystal field generated by the surface,[5, 6] leaving the spin $\mathbf{S}$ as the only viable parameter for control. Even in the case of a partially preserved orbital moment, the spin-orbit interaction can act to create superpositions of the orbital angular momentum and spin states, meaning that only the total momentum $\mathbf{L} + \mathbf{S}$ is preserved. In that case, independent excitations of $\mathbf{L}$ and $\mathbf{S}$ cannot occur.

Quenching of the orbital angular momentum directly affects the stability and lifetime of the atom's magnetization.[7, 8] The viability of information processing applications based on single atoms is, however, contingent on the spin stably maintaining its direction, and thus its magnetization, over time—which necessitates a large single-site magnetic anisotropy, as well as a slow relaxation of the magnetization. The energy barrier to flip the magnetic moment is determined by the magnetic anisotropy energy (MAE), which arises from the interplay between the crystal field and spin-orbit coupling. Specifically, the Coulomb potential generated by the crystal breaks the spherical symmetry of the free atom, thereby lending the orbital moment a certain orientation with respect to the crystallographic axes.[5] However, in the case of an almost fully quenched $\mathbf{L}$, the spin-orbit coupling only acts to higher order to produce single-site magnetic anisotropy, which leads to MAE values far below the atomic spin-orbit coupling strength. Consequently, the crystal symmetry at the atomic site—and the overlap of the atomic orbitals with the surrounding ligands—plays a crucial role in preserving the orbital angular momentum of the atom and enhancing the MAE.

Engineering the local environment of the single atom to produce an axial crystal field can have significant consequences on preserving the free-atom orbital moment, and consequently, increasing the magnetic anisotropy.[7, 8] $3d$ transition elements are particularly appealing as the magnetic atoms of choice, as, in addition to their natural abundance, they can be easily deposited on surfaces and probed locally by scanning tunneling microscopy (STM) and spectroscopy. This is illustrated by STM experiments performed on Fe and Co atoms bound to the oxygen site of the MgO/Ag(100) surface, where the local symmetry ensures a nearly axial crystal field. The resultant orbital moment–which is nearly preserved in the out-of-plane direction for the Fe atoms, and fully preserved for the Co atoms— gives rise to large zero-field splittings of, respectively, 14 meV[7] and 58 meV[8]. However, in both of these cases, the energy multiplets evolve under the crystal field and spin-orbit coupling to become a mixture of S and L states, and accordingly, the transitions probed by inelastic tunneling spectroscopy (IETS) show that variations in $\mathbf{L}$ are associated with variations in $\mathbf{S}$.

In this work, we present a single-atom spin system that combines a large MAE with an orbital angular moment that remains fully unquenched along the uniaxial direction. This situation is realized by placing Fe atoms atop the fourfold symmetric nitrogen binding site of the $Cu_2N/Cu_3Au(100)$ surface, thus engendering a zero-field splitting of 18 meV. We demonstrate that we are able to fully rotate the preserved orbital moment via a single-electron process between the tip and atom, without altering the spin state of the atom. Alternatively, we observe a distinct spin excitation, which does not affect the orbital moment. These finding are understood in terms of first-principles density functional theory (DFT) and electronic multiplet calculations.

## Results

### Origin of the unquenched orbital moment

The $Cu_2N$ surface,[9] in addition to providing protection to magnetic moments from electronic scattering, enables reliable and large-scale atom manipulation.[10–15] Fe atoms on the $Cu_2N$ lattice preferentially bind to the Cu-site, where the local $C_{2v}$ symmetry produces a partially unquenched orbital moment resulting in in-plane uniaxial magnetic anisotropy energies of $\sim 5$ meV.[11, 14] A higher symmetry can be achieved, however, by coordinating the Fe atom atop the N-site instead, which, in principle, could preserve the orbital moment even more, and thus lead to larger anisotropy values. N-site adsorption on $Cu_2N$ is also preferable over Cu-site adsorption, in that placing an Fe atom on an N-site requires one less atom manipulation procedure,[16] vastly improving possibilities for building extended spin arrays. However, previous studies reported that no spin-flip excitations could be resolved for Fe atoms bound to N-sites.[9]

We use a low-temperature STM to perform controlled single-atom manipulation and inelastic tunneling spectroscopy. We coordinate Fe atoms, deposited on an insulating layer of $Cu_2N$ that is grown on a $Cu_3Au(100)$ substrate,[17] atop the N and Cu sites of the lattice (Fig. 1a). The apparent height of the Fe atoms atop the N-sites is $\sim 3.1$ Å, roughly 0.4 Å higher than those on Cu sites. The N binding site is fourfold symmetric ($C_{4v}$), with four Cu atoms as nearest neighbours, a lateral distance of 1.77 Å away (Fig.1b). DFT calculations (see Supplementary Note 1 for additional information) indicate that the N atom atop which the magnetic atom is bound is displaced upwards by 0.3 Å with respect to the pristine surface configuration. The calculated magnetic moment for the spin of the Fe atom, considering an on-site Coulomb interactions U = 5 eV, is $\mu_S \approx 4.36\ \mu_B$, with $\mu_B$ the Bohr magneton; this indicates a local spin $S = 2$. The DFT-calculated valence electron spin density (Fig.1c) shows that the axial symmetry is largely intact. Thus, we can expect the orbital moment to be preserved in the out-of-plane direction, while it is quenched in-plane. The typical overestimation of the orbital momentum quenching by DFT calculations precludes a quantitative description of **L**, and thus, of the resulting MAE.[18, 19]

Instead, here we adopt an alternative strategy: we carry out an electronic multiplet calculation based on a point-charge model (PCM) description of the crystal field, where electron-electron repulsion between Fe $d$-electrons, spin-orbit coupling, and Zeeman contributions are considered explicitly (Supplementary Note 2).[20, 21] The atomic positions and charges are extracted from the DFT calculations. A similar method was applied successfully to study the spin excitations of Fe on MgO.[7]

**Describing the electronic multiplet**

The lowest energy levels derived from the multiplet calculations are shown in Fig. 2a. The crystal field (CF) contribution is separated into its axial and transverse components: the former splits off a tenfold ground state degeneracy, while the latter splits this into two spin quintuplets. The spin-orbit coupling—where we used $\lambda = -9.60$ meV for the PCM, and $-9.41$ meV for the spin-orbit model—partially lifts the degeneracy within the two quintuplets. Finally, the magnetic field $B_z$ along the out-of-plane direction breaks all remaining degeneracies. At a non-zero field in the out-of-plane direction, the lowest two states have orbital moments $L_z = \pm 1.98$, closely approaching the free-atom value. Below, we will approximate these two states as $L_z = \pm 2$. Notably, the multiplets evolve under the crystal field and spin-orbit coupling to become nearly pure product states of the $S_z$, $L_z$ eigenstates. This separation of the spin and orbital degrees of freedom is permitted by the relative dominance of the magnetic anisotropy energy over the strength of the spin-orbit coupling. In fact, use of the total angular momentum basis is not adequate here, since the magnetic anisotropy terms do not commute with the total angular momentum ($\hat{\mathbf{J}}^2$ and $\hat{\mathbf{J}}_z$).

When interpreting spin excitation spectroscopy on individual magnetic atoms, it is convenient to employ an effective spin Hamiltonian.[11, 12, 22] However, in this situation the unquenched orbital moment makes the effective spin framework incomplete.[23] Instead, we use the following anisotropic spin-orbit Hamiltonian[23]

$$\hat{\mathcal{H}} = B_2^0\hat{O}_2^0 + B_4^0\hat{O}_4^0 + B_4^4\hat{O}_4^4 + \lambda_{SO}\hat{\mathbf{L}}\cdot\hat{\mathbf{S}} + \mu_B(\hat{\mathbf{L}} + 2\hat{\mathbf{S}})\cdot\mathbf{B},\tag{1}$$

where $\hat{O}_k^q$ are the Stevens operators, which in this case are applied to the eigenstates of the orbital moment, and $B_p^q$ are their associated coefficients, respectively. The last term represents the Zeeman energy due to an external field $\mathbf{B}$. As we consider both the spin $\mathbf{S}$ and orbital moment $\mathbf{L}$, there is no need to invoke the Landé $g$-factor. The results of this model, implemented with optimal fitting parameters (see Supplementary Note 3), are also depicted in Fig. 2a. Note that there is perfect agreement between the PCM and the spin-orbit model presented in Eq. (1). We additionally confirm these results using electronic multiplet calculations derived using the Wannier Hamiltonian to approximate the crystal and ligand fields produced by the surface atoms (see Supplementary Note 5). This approach provides a more accurate quantitative description, and additionally accounts for charge transfer and surface polarization.

**Independent spin and orbital excitations**

We perform an IETS measurement with an out-of-plane field of 4 T, revealing a splitting of the zero-field spin excitation, with threshold voltages at 17.9±0.7 meV and 19.4±0.7 meV, as shown in Fig. 2b. These transitions can only be probed with a tip that is functionalized by picking up individual Fe atoms from the surface. The results of Fig. 2a allow us to uniquely assign the observed transitions to excitations between specific states. When describing these states, we choose to use product state notation since $S_z$ and $L_z$ are approximately good quantum numbers here. The lower energy excitations are spin-only transitions ($\Delta S_z = \pm 1$, $\Delta L_z = 0$) corresponding to an excitation from the ground state $|S_z\rangle |L_z\rangle = |-2\rangle |-2\rangle \equiv |0\rangle$ to $|-1\rangle |-2\rangle \equiv |2\rangle$, corresponding to an excitation threshold voltage $V_{02}$, as well as from the $|+2\rangle |+2\rangle \equiv |1\rangle$ state to $|+1\rangle |+2\rangle \equiv |3\rangle$, with threshold $V_{13}$ (Fig. 2d). At zero field, $|V_{02}| = |V_{13}| = 18.4\pm0.6$ meV.

In addition, we observe a higher energy excitation at 73.9±0.8 meV (see Fig. 2b), which we denote by the threshold voltage $V_{08}$. This feature corresponds to an excitation from the ground state $|0\rangle$ to the excited state $|-2\rangle |+2\rangle \equiv |8\rangle$; i.e., going from the lower spin quintuplet to the upper spin quintuplet (see Fig. 2d). A detailed analysis of the calculated transition strengths (see Supplementary Note 6) confirms that an excitation from $|0\rangle \rightarrow |8\rangle$ occurs with a much larger amplitude than from other possible paths, such as transitions from $|0\rangle \rightarrow |6\rangle$ or $|2\rangle \rightarrow |8\rangle$. Additionally, the energy at which this transition occurs quantitatively agrees with the energy difference between the states $|0\rangle$ and $|8\rangle$ across the various models we implement, namely the point-charge and Wannier models (Fig. 2b and Supplementary Fig. 3).

Unlike a conventional spin excitation—in which the tunneling electron spin only interacts with the atom's spin ($|\Delta S_z| \leq 1$), leaving the orbital moment unchanged—we observe an independent excitation of only the orbital moment, with $\Delta L_z = 4$. Although orbital excitations have been previously reported,[24, 25] here we observe a full, independent rotation of an unquenched orbital moment. These transitions are not accounted for by the usual spin exchange terms $J\mathbf{S} \cdot \boldsymbol{\sigma}$,[26, 27] even when the orbital and spin degrees of freedom are accounted for, as in Eq. (1).

Rather, this orbital transition can be understood via a co-tunneling path that takes into account both the spin and the orbital momentum of the initial, intermediate and final states, as depicted in Fig. 2c.[28, 29] Since the transition is expected to occur with similar amplitude for the hole and electron charged states, we will focus on the latter for the following discussion. In this case, the dominant channel is mediated through the negatively charged intermediate state $|S_z = -3/2\rangle |L_z = 0\rangle$. Accordingly, the co-tunneling transition amplitude between the ground state $|0\rangle$ and the excited state $|8\rangle$ can be understood by introducing the creation and annihilation operators, $\hat{d}^{\dagger}_{\sigma_z \ell_z}$ and $\hat{d}_{\sigma_z \ell_z}$, for an electron with spin $\sigma_z$ in an orbital with angular momentum $\ell_z$ (centered on the atom). The dominant transition amplitude between states $|0\rangle$ and $|8\rangle$ is thus proportional to[28, 29]

$$\langle +2| \langle -2| \hat{d}_{+\frac{1}{2}, -2} |-3/2\rangle |0\rangle \langle 0| \langle -3/2| \hat{d}^{\dagger}_{+\frac{1}{2}, +2} |-2\rangle |-2\rangle. \tag{2}$$

This co-tunneling path corresponds to a spin-up electron tunneling onto the $\ell_z = +2$ orbital, thus creating a charged virtual state with a net spin $S_z = -3/2$ and orbital moment $L_z = 0$. An electron then tunnels off the $\ell_z = -2$ orbital, restoring the net spin to $S_z = -2$ and changing the orbital moment to $L_z = +2$, thereby completing the $\Delta S_z = 0$, $\Delta L_z = 4$ transition. Thus, we show independent transitions of the spin and unquenched orbital angular momenta, where we can rotate one of these atomic degrees of freedom without affecting the other.

At first sight, a $\Delta L_z = 4$ transition may seem to violate conservation of total angular momentum. However, we point out that the orbital moment of a freely propagating electron is defined relative to an arbitrary origin, and can therefore, unlike the spin, assume an arbitrary value. An electron tunneling from the tip is thus free to carry an orbital moment, and inelastically excite the atomic orbital moment. Within this framework, conservation of total momentum can be understood in terms of the Einstein-de Haas effect, wherein the angular momentum of the tunnelling electron is translated into an infinitesimal rotation of the macroscopic lattice.[30, 31]

We trace the evolution of the magnetic behavior of the single atom as a function of external field: in Fig. 3a, b, and c we show IETS measurements of the spin and orbital excitations, performed for a range of discrete fields up to 5 T. In both cases, we observe the Zeeman effect as a shift towards higher threshold voltages at higher field. The measurements indicate a shift in the threshold voltage of 0.23±0.04 meV/T and 0.31±0.05 meV/T for the spin and orbital transitions, respectively (Fig. 3d). When expressed in terms of an effective $S = 2$ spin model in the absence of orbital angular momentum,[11] the shift for the spin excitation would correspond to a Landé factor of ~ 3, on par with previously reported large values.[7, 8, 32]

Additionally, we expect the orbital excitation to correspond to two transitions: $|0\rangle \rightarrow |8\rangle$ and $|1\rangle \rightarrow |9\rangle$, which should split as a function of magnetic field due to the Zeeman effect. We observe that the step is broadened as the field is increased, which is compatible with a splitting of $V_{08}$ and $V_{19}$. We note that $V_{19}$ is marked by a step down in the differential conductance, which is due to spin-polarized elastic conductance, combined with a reconfiguration of the occupation of states $|0\rangle$ and $|1\rangle$ around the threshold voltages.

The observed behavior is well reproduced by the transport calculations derived from the point-charge model. In fact, the high degree of agreement between the experimentally and theoretically derived results here is remarkable, as the point-charge calculations are based solely on DFT results, and thereby don't have any additional fitting parameters, except for a screening factor applied to the free-atom spin-orbit coupling (adjusted only to reproduce the energy of the spin excitation). However, the threshold voltage corresponding to the orbital excitation is off by $\sim 1.4$ meV when comparing the transport calculations to the experimental data. In order to properly compare the evolution of the step, we correct for this shift in Fig. 3c. We note that the measurements shown in panels a and b are obtained on different atoms, using a different functionalized tip, than measurements shown in c and d—this causes a slight offset in the measured threshold voltages, presumably due to the tip field or variations in the local environment. We try to account for these variations, and the ambiguity in defining the threshold energy due to the unusual lineshape of the spin excitations, in the error associated with $V_{02}$, $V_{13}$, and $V_{08}$.

The field dependence of the threshold voltages confirm our assignation of the observed transitions to those belonging to independent excitations of the spin and orbital momentum. The ratio between the rate of change of the $V_{08}$ and $V_{02}$ transitions, amongst the various models we implement, is consistently between 1.6 to 2 (refer to Supplementary Note 7); experimentally, we observe a ratio of 1.3±0.3. In contrast, the $V_{06}$ and $V_{17}$ transitions, which corresponds to full rotations of the orbital moment along with a partial rotation of the spin, are expected to shift much faster under the effect of external field, with a rate of change 3 times that of $V_{02}$.

In the absence of non-equilibrium effects, inelastic spin excitations ($\Delta S_z = \pm 1$) are characterized by approximately square steps in the differential conductance,[33] which originate from co-tunneling events.[26, 27] However, additional nonlinearities may appear at the threshold voltage due to changes in the instantaneous spin state of the atom, which modify the magnetoresistance of the junction, and thus, the d$I$/d$V$ lineshapes.[13, 34] The dynamical effects that we observe at the inelastic tunneling threshold voltage for the spin excitation (Fig. 4a) are indicative of relaxation times from state $|1\rangle$ longer than the average time between two tunneling electrons (∼200 ps at 1 nA).

As the presence of non-equilibrium features is attributed to dynamic processes linked to the inelastic electron transport, they are expected to be conductance-dependent. We investigate this dependence by performing d$I$/d$V$ measurements as a function of current set-point, as shown in Fig. 4a and b. For this range of conductance values, we observe a decrease in the strength of the nonlinearity with increasing tunnel current[34, 35] and a shift in the inelastic steps, both of which are due to the local field from the exchange interaction between the Fe atom and the tip.[36]

Further insight can be obtained by simulating the non-equilibrium dynamics of the local spin (Fig. 4c). This is done on two fronts: on one hand, starting from the point-charge model calculation, we calculate the transition rates and the non-equilibrium occupations in the weak coupling limit using a co-tunneling description of transport.[28, 29] On the other, we use the spin-orbit model Eq. (1) exchange coupled to the itinerant electrons. In both cases, the evolution of the occupation is accounted for by a Pauli master equation.[26, 27] Tracing the occupation of the two lowest spin states as a function of voltage (Fig. 4d) delineates that below the inelastic threshold voltage, the ground state occupation exceeds 90%. Once the applied voltage reaches the excitation threshold, spin-flip excitations cause a significant drop in the occupation of $|0\rangle$.

**Discussion**

By coordinating a magnetic atom atop the fourfold symmetric nitrogen binding of the $Cu_2N$ lattice, we have realized a single atom system with a large magnetic anisotropy, which follows from a preserved orbital angular momentum, an ingredient that is essential to the application of magnetic atoms in magnetic storage and information processing. In this system, under the effects of the crystal field and spin-orbit coupling, the multiplets emerge as nearly pure **L** and **S** product states, which allows us to treat these parameters as two independent degrees of freedom. We demonstrate independent control over both the spin and orbital moment, showing a full inversion of the orbital moment by means of a single electron, without affecting the spin.

As control over the orbital angular momentum shows many parallels to that of the spin momentum, we believe that this development adds a new dimension to studies on single-atom magnetism. Moreover, as Fe atoms bound to N-sites are easily manipulable, these results form a promising basis for future research on extended lattices, that can interact through both the spin and orbital angular momentum.

**Methods**

**Experimental Considerations**

The experiment was conducted using a commercial low-temperature, high magnetic field STM (Unisoku USM1300s). The sample was prepared *in situ*: monolayer insulating islands of $Cu_2N$ were grown on a $Cu_3Au(100)$ surface via nitrogen sputtering, and subsequently single Fe atoms were evaporated on the cold sample using electron-beam physical vapor deposition. A variable out-of-plane magnetic field was applied using a superconducting magnetic coil. IETS measurements were performed using standard lock-in detection techniques at base operating temperature (330 mK).

**Multiplet Calculations for Fe/$Cu_2N$/$Cu_3Au(100)$ system**

For the multiplet calculations of Fe atoms, we used an archetypal value of the Hubbard repulsion $U = 5.208$ eV ($U - J = 5$ eV).[37, 38] We have taken the atomic values of $\langle r^2 \rangle = 1.393$ and $\langle r^4 \rangle = 4.496$ atomic units.[23] Instead of correcting the $\langle r^2 \rangle$ and $\langle r^4 \rangle$ parameters due to covalency and other known limitations of the point charges, we have taken the spin-orbit coupling $\lambda$ as a fitting parameter to reproduce the 18 meV step. The optimal fitting is found when the spin-orbit coupling is screened by a factor 0.738, which translates into a (many-body) effective spin orbit coupling of –9.60 meV. The transport calculations under the co-tunneling regime were carried out assuming electron-hole symmetry, i.e., $E_{0-} - E_0 - E_F = E_F - E_0 + E_{0+}$. For the surface hybridization constants, we take $V_{k_F, S} = 0.562$ eV, and for the tip-hybridization $V_{k_F, d_{z^2}} = 0.183$ eV $= 6V_{k_F, d_{x^2-y^2}} = 6V_{k_F, d_{xy}}$.

**Parameters of the anisotropic spin-orbit Hamiltonian**

The parameters $B_p^q$ and $\lambda_{SO}$ of the spin-orbit Hamiltonian (Eq. (1)) were obtained by fitting the corresponding energy spectrum to the results of the multiorbital electronic multiplet Hamiltonian at zero magnetic field. The best fit was obtained for $B_2^0 = -1.404$ eV, $B_4^0 = 0.188$ eV, and $B_4^4 = 0.16$ meV, which indicates an almost pure uniaxial easy axis system. The value obtained for the spin-orbit coupling is, $\lambda_{SO} = -9.41$ meV. Details regarding the fitting procedure are delineated in Supplementary Note 3. Additionally, the coupling to the surface was taken to be $(\rho J_{K,S})=0.25$, where $\rho$ is the density of states at the Fermi energy and $J_{K,S}$ is the Kondo exchange coupling with the surface, while $(\rho J_{K,T})=0.0484$ for the tip. In addition, a direct tunnelling term of $(\rho T)=0.25$ was also assumed (we have assumed the same density of states for the surface and the tip).

**Data Availability**

All data presented in this work is publicly available with identifier (DOI) 10.5281/zenodo.3959042

**Acknowledgments**

The authors thank the Netherlands Organisation for Scientific Research (NWO) and the European Research Council (ERC Starting Grant 676895 'SPINCAD'). F.D. acknowledges financial support from Basque Government, grant IT986-16 and Canary

Islands program Viera y Clavijo (Ref. 2017/0000231). J.W.G. acknowledges financial support from FONDECYT: Iniciación en Investigación 2019 grant N. 11190934 (Chile).

**Competing Interests:**

The authors declare no competing interests.

**Author Contributions**

R.R., D.C. and A.F.O. conceived the experiment, R.R., D.C., and J.G. acquired the data. R.R. analyzed the data. J.W.G. and F.D. performed theoretical calculations. R.R., F.D., and A.F.O. wrote the manuscript, with considerable contributions from all listed authors. A.F.O. supervised the experimental work.

chains. *Phys. Rev. Lett.* **119**, 217201 (2017).

**Figure Legends**

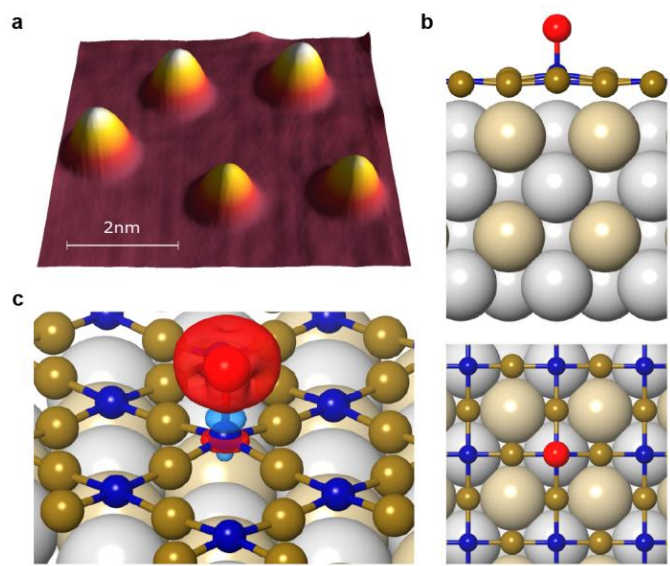

**Fig. 1.** Fe atoms atop the nitrogen site of the Cu₂N surface. **a** STM constant-current topography (30 mV, 20 pA) of Fe atoms on a Cu₂N/Cu₃Au(100) surface. To the bottom right, there are two Fe atoms bound to copper sites, and at the top, three Fe atoms atop nitrogen sites with larger apparent heights. Scale bar: 2 nm. **b** Side and top view of the binding geometry for the Fe atom (red) atop a N atom in the Cu₂N

network (Cu brown, N blue) on a $Cu_3Au$ crystal (Cu grey, Au yellow). **c** Calculated positive (red) and negative (blue) electron spin density.

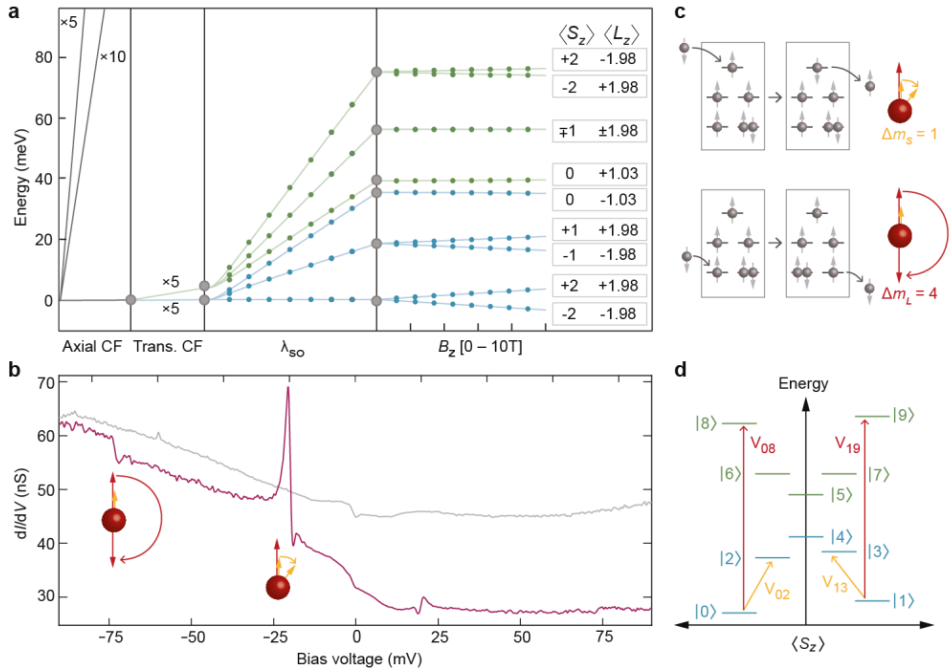

**Fig. 2.** Energy spectra and IETS measurements. **a** Energy spectra derived using both the PCM (solid lines) and the spin-orbit model (dots). The expectation values of $S_z$ and $L_z$ are indicated for each state. The transverse crystal field generates two distinct spin quintuplets (blue and green). The energy scale is defined relative to the ground state energy, except in the rightmost panel where the Zeeman splitting is considered, in which case the absolute energies are plotted. **b** Differential conductance ($dI/dV$) spectroscopy performed with a functionalized tip on a single Fe atom (magenta) and on bare $Cu_2N$ (gray) ($T = 0.3$ K, $B_z = 4$ T, 400 μV modulation, taken at -90 mV, 8 nA). **c** Co-tunneling mechanism for inelastic excitations of the spin (top) and orbital (bottom) momenta. Each rectangle represents the energy levels of the five $\ell_z$ orbitals as follows: $\ell_z = \pm 2$ (bottom), $\ell_z = \pm 1$ (middle), $\ell_z = 0$ (top). In the case of a spin-excitation, the electrons are free to tunnel on and off the same singly-occupied orbital. **d**

Schematic representation of the two lowest quintuplets, with the spin and orbital transitions probed by IETS marked with arrows.

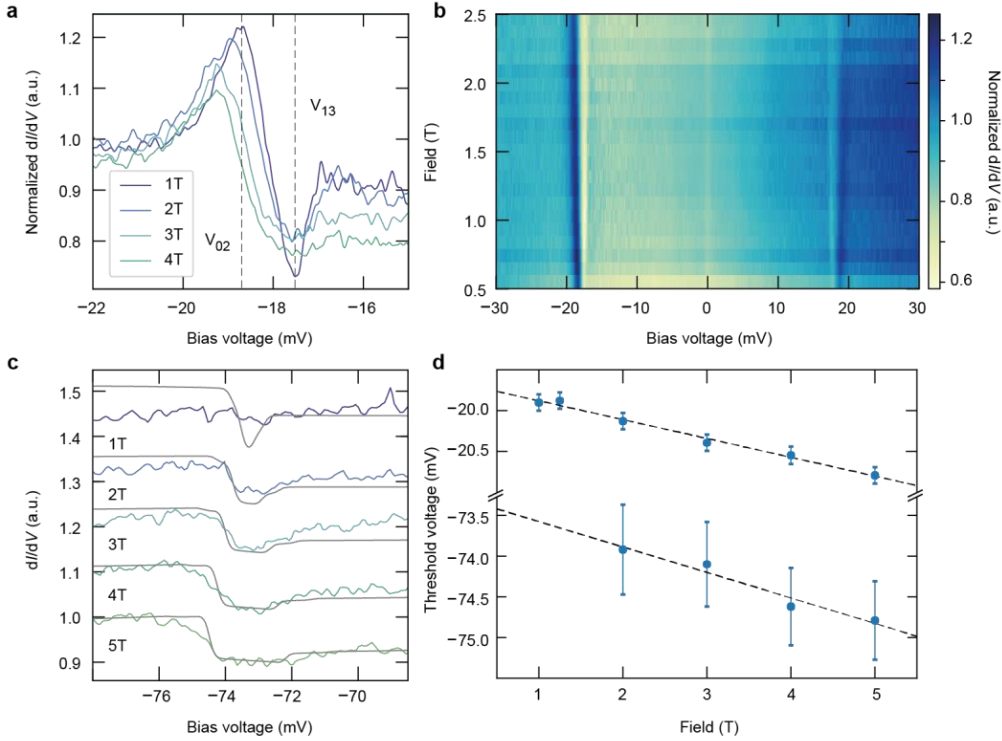

**Fig. 3.** Magnetic field dependence of the spin and orbital excitations. **a** Differential conductance spectroscopy for different values of the external magnetic field, with the dotted lines denoting threshold voltages $V_{02}$ and $V_{13}$ at 1 T. **b** Color map of d$I$/d$V$ spectroscopy as a function of magnetic field. **c** Differential conductance spectroscopy (conductance set-point of -90 mV, 8 nA) showing a transition at ~74 meV, for various magnetic fields, normalized and shifted vertically (with respect to the 5 T spectrum) for clarity. Overlaid are the corresponding transport calculations (grey) derived from the point charge model, horizontally shifted by -1.4 meV to match the experimentally derived threshold voltage. **d** The measured threshold voltages $V_{02}$ and $V_{08}$ as a function of the external magnetic field. The error bars here only account for the uncertainty in the fit of the step position.

Dashed lines are linear fits, indicating a shift of 0.23±0.04 meV/T for the $V_{02}$ transition and 0.31±0.05 meV/T for the $V_{08}$ transition.

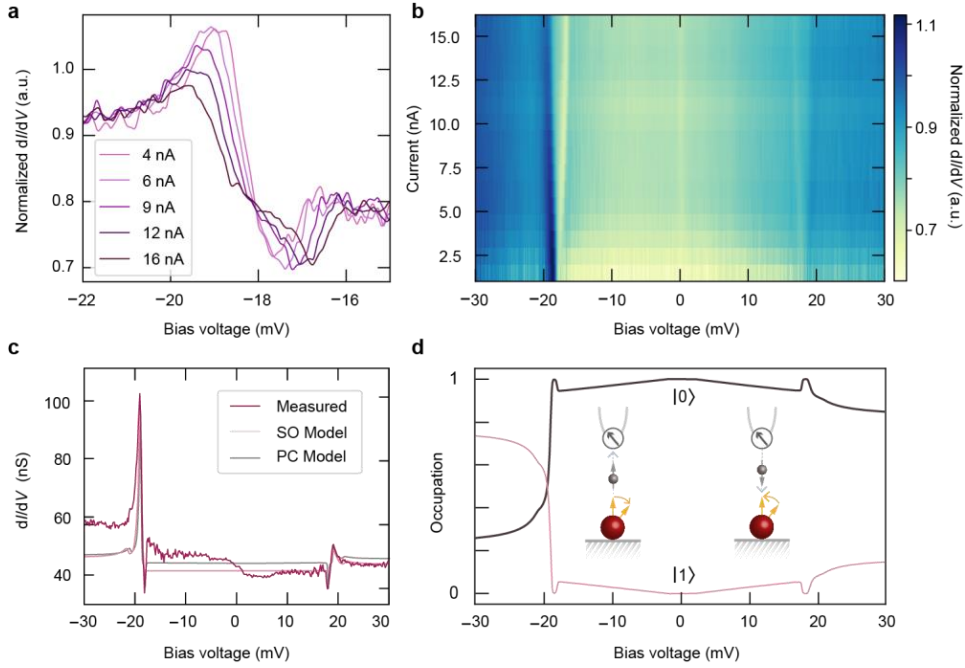

**Fig. 4.** Conductance dependence of non-equilibrium electron transport. **a** Differential conductance spectroscopy for different conductance values. **b** Color map of conductance-dependent $dI/dV$ spectroscopy. The same experimental parameters ($B_z = 4$ T, $T = 0.3$ K, 150 µV modulation) apply for both **a** and **b**. **c** Spectroscopy measurement at $B_z = 2$ T (magenta), taken at a conductance set-point of -90 mV, 8 nA, compared to normalized transport calculations derived from the point-charge (grey) and spin-orbit (pink) models. **d** Calculated voltage-dependent occupation of the two lowest energy states using the point-charge model ($P_T = -0.3$).